\documentclass[aps,prl,preprint,superscriptaddress]{revtex4-1}
\usepackage{hyperref}
\usepackage{graphicx}
\usepackage{amsmath,mathrsfs}
\usepackage{amsfonts}
\usepackage{amssymb}
\usepackage{bm}
\usepackage{xcolor}

\newcommand\J{\Psi}
\renewcommand\d{\partial}
\newcommand{\no}{\nonumber}

\begin{document}

\title{On Berezinskii-Kosterlitz-Thouless Phase Transition in Quasi-One 
       Dimensional Bose-Einstein Condensate}

\author{Vivek M. Vyas}
\affiliation{Indian Institute of Science Education and Research (IISER)- Kolkata,
 Mohanpur, Nadia-741252, India}
\email{vivek@iiserkol.ac.in}

\author{Sandeep Gautam}
\affiliation{Physical Research Laboratory, Navrangpura, Ahmedabad-380009, India}
\email{sandeep@prl.res.in}

\author{Prasanta K. Panigrahi}
\affiliation{Indian Institute of Science Education and Research (IISER)- Kolkata,
 Mohanpur, Nadia-741252, India}

\date{\today}

\begin{abstract}
We show that quasi-one dimensional Bose-Einstein condensate under suitable conditions can exhibit a  
Berezinskii-Kosterlitz-Thouless phase transition. The role played by quantized vortices in the two dimensional case, is played in this case by dark solitons. We find that the critical temperature for this transition lies in nano Kelvin range and below, for a wide range of experimentally accessible parameters. It is seen that the high temperature (disordered) phase differs from the low temperature  (ordered) phase in terms of phase coherence, which can be used as an experimental signature for observing this transition.   
\end{abstract}

\maketitle

The continuum version of 2D XY model appears in diverse areas of physics 
ranging from spin systems to superfluids \cite{Chaikin,Huang,Mackenzie1,Mackenzie2}, with its Lagrangian being given by
$\mathcal{L} = - \rho ( \vec{\nabla} \theta )^{2}.$
It is well known that this model exhibits singular vortex solutions: $\roarrow{\nabla}\theta \propto \frac{\hat{\phi}}{r}$ (in polar coordinates) \cite{Chaikin,Huang}. Identifying two dimensional Cartesian plane 
with complex plane, allows one to write vortex solution in a rather elegant form 
\cite{Nagaosa}:
\begin{align}
\theta(x,y) = \theta(z=x+i y) = \pm \text{Im} \, \text{ln}(z-z_{0}),
\end{align}
where $z_{0}$ is the location of vortex. It is evident that the above solution exhibits 
a discontinuous jump as one traverses along a closed loop enclosing $z_{0}$. 
This multivaluedness or discontinuity is a characteristic feature of these 
vortices and distinguishes them from other solutions. It has been shown that 
the Helmholtz free energy of a system of $K$ vortices is given by \cite{Chaikin,Huang},
\begin{align}
F = K \left( 2 \pi \rho - 2 k T \right) \text{ln} \left(\frac{L}{b} 
\right),
\end{align}
where $k$ stands for Boltzmann constant, $L$ represents system size, and $b$ is vortex core radius.
Hence, one finds that below critical temperature $T_{BKT}= \frac{\pi \rho}{k}$, free energy is minimized by having $K=0$, which means that at temperature low enough vortices are thermodynamically 
unstable. However, when $T>T_{BKT}$, free energy minimization requires
having $K$ as large as possible. So the critical temperature $T_{BKT}$ marks the occurrence of spontaneous proliferation of vortices, which makes condensate inhomogeneous. This is the
celebrated Berezinskii-Kosterlitz-Thouless (BKT) phase transition \cite{Berezinskii,Kosterlitz}. 
This phase transition was found to be of infinite order, and unlike second order phase transition, here the ordered (low temperature) 
phase differs from disordered (high temperature) phase not in terms of 
symmetry of ground state but rather in terms of topology of the ground state 
\cite{Chaikin,Huang}. In the low temperature phase, 
system exhibits a quasi long range order, where 
$\langle e^{i \theta (x)} e^{i \theta (y)} \rangle \approx |x - y|^{-\alpha}$
for some constant $\alpha$. Above  
critical temperature the system has vortex excitations, which cause rapid phase variation,
and one sees that correlation function 
$\langle e^{i \theta (x)} e^{i \theta (y)} 
\rangle \approx e^{- \beta |x - y|}$ ($\beta$ is a constant).
This means that the quasi long range order present at low temperature is destroyed due to 
vortices, resulting in rapid decay in correlation.

It is established by now that in the mean field approximation, physics of Bose-Einstein condensate (BEC) is well described by the Gross-Pitaevskii (GP) equation \cite{Pethick}:
\begin{equation}
\left( - \frac{\hbar^2}{2m} \roarrow{\nabla}^{2} + V(\mathbf r) + U_0|\Psi|^2\right)
 \Psi(\mathbf r, t) = i\hbar\frac{\partial \Psi(\mathbf r, t)}{\partial t}. 
\label{gp.eq}
\end{equation}
Here $V(\mathbf r)$ is the external trapping potential, and 
$U_0 = 4\pi\hbar^2 a/m$ is the coupling constant for inter-atomic interaction, where $m$ and $a$ are atomic mass and $s$-wave scattering length, respectively. The above partial differential equation is an 
equation of motion describing space-time dependence of BEC, which itself is a 
ground state of a system of interacting bosons. In case when trapping potential is tightly confining along Z-direction and absent (or very weak) along the other two spatial dimensions, 
dynamics of BEC occurs only along the latter two spatial dimensions. The condensate function can then be written as $\J(\mathbf{r},t)= f(z) \phi(x,y,t)$, where $f(z)$ is a localized Gaussian wave 
packet along Z-direction, and $\phi(x,y,t)$ describes condensate profile in the
XY plane. Equation governing dynamics of such quasi-2D
BEC can be obtained by integrating out degrees of freedom along Z-direction. Condensate density, in absence of any potential acting along XY directions, can be expected to be stationary 
(temporally) and spatially uniform: $|\phi(x,y,t)|^{2}=\rho$, 
where $\rho=\text{constant}$, and $\phi$ field can be written in terms of the polar variables: 
$\phi = \sqrt{\rho} e^{i \theta(r,t)}$. It can be shown that in static case, Lagrangian for quasi-2D BEC reads as
$\mathscr{L}_{2D} \propto - \rho ( \roarrow{\nabla} 
\theta )^{2}$, 
which exactly matches with that of 2D XY model. Hence, one finds that dynamics of quasi-2D BEC is actually captured by XY model, and so one expects to observe BKT transition in this system.

Above discussion is oversimplified, and in reality, bosons are trapped not 
only in Z-direction but also in XY directions. Because of this, there appears
a confining potential defined over XY plane. This makes the system 
inhomogeneous, and hence translational invariance is lost. In the last few years, 
there have been significant activity, on both theoretical and experimental 
fronts, to gain more understanding about nature of BKT transition in these 
systems \cite{Schweikhard,Simula-05,Simula-06}. A few years back, 
Hadzibabic \emph{et al.,} reported a direct observation of BKT transition in 
BEC \cite{Hadzibabic}. Further, in recent years study of dynamics of vortices 
in BEC systems especially, vortex dipoles (vortex-antivortex pairs) has 
received significant attention \cite{Susanto,Rodrigues,Gladush,Sasaki,
Neely,Freilich}. Vortex dipoles are also produced as the decay products of 
dark solitons in 2D BECs due to onset of the snake instability 
\cite{Theocharis}.

In light of above discussion, it is natural to ask, if a BKT phase 
transition occurs in BEC in spatial dimensions other than two. The answer 
is crucially dependent on another question: Do the counterparts of vortices
exist in dimensions other than two ? Analogues of vortices in three dimension 
are well known, they are vortex tubes or rings \cite{Anderson,Dutton}. 
Being spatially extended objects, interaction between them is complicated, which makes the 
problem rather nontrivial \cite{Shenoy,Kleinert}. Instead, we ask whether 
there are any analogues of vortices in one dimension? If yes, then does one 
dimensional BEC display BKT type phase transition? Answer to the first 
question is known. The so called dark soliton solutions are known to be 
analogues of vortices in one dimension \cite{Carr,Frantzeskakis}. In what 
follows, we answer the second question affirmatively. After introducing 
dark soliton solution of GP equation, we indicate its analogy with two 
dimensional vortex. Thereafter, we show using exactly similar free energy
argument as was originally used by Kosterlitz and Thouless \cite{Kosterlitz} 
(which is also used in the above discussion) that it is indeed possible to have 
a finite temperature BKT phase transition in quasi-one dimensional BEC. We 
estimate the critical temperature for the same and find that it can be tuned 
over a broad range by changing scattering length and number density. 
Subsequently, we also show that the long range phase coherence exhibited by 
the system at low temperature vanishes at the temperatures above critical 
temperature. We conclude by mentioning that it may be possible to 
experimentally realize this phase transition.

As mentioned earlier, GP equation (\ref{gp.eq}) governs motion of order parameter 
$\Psi(\mathbf r,t)$, which satisfies normalization condition
$ \int|\Psi(\mathbf r,t)|^2d{\mathbf r} = N$,
where $N$ is the number of the bosons. In present case, we consider trapping potential  
to be harmonic along the radial direction and infinite square 
well along the axial direction, \emph{i.e.,} 
\begin{equation}
  V(r,z)=\left \{ \begin{aligned}
          &\frac{m}{2}\omega^2 r^2   
                  && \text{if } -\frac{L}{2}< z <\frac{L}{2}, \\
          &\infty      &&\text{if }   |z| \ge {\frac{L}{2}},
       \end{aligned} \right .
\label{triple_pot}              
\end{equation}
where $\omega$ is the radial trap frequency and $L$ is the size of infinite
square well potential along axial direction. The infinite square well potential
can be created by applying two blue detuned laser beams with narrow beam 
waist and sufficiently high intensity at $|z| = L/2$.

Since the potential becomes infinitely high at and after $|z| = L/2$, 
condensate density vanishes at $|z| = L/2$ and increases smoothly as one moves 
towards the center. The characteristic length scale over which BEC recovers 
its bulk density $n_0$ is called healing length or coherence length 
$\xi = \frac{1}{\sqrt{8\pi n_0 a}}$.
The characteristic length of the BEC along radial direction is oscillator 
length $a_{\rm osc}=\sqrt{\hbar/(m\omega)}$. 
When $a_{\rm osc}\lesssim\xi$, $a_{\rm osc}\ll L$, and BEC is in the weakly 
interacting regime $2\pi a N \int r|\Psi(r,z)|^2 dr\ll1$, the order 
parameter can be factorized into radial and axial parts \cite{Salasnich}
\begin{equation}
  \Psi(r,z,t) = \chi(r)\psi(z,t),
\label{eq.psi}
\end{equation}
where $\chi(r)$ is the normalized ground state of radial trapping potential, 
which can be considered to be time independent 
provided, $kT \ll \hbar\omega$. We consider $\chi(r)$ to be normalized to
unity $(2\pi\int r|\chi(r)|^2dr =1)$; then from Eq.(\ref{gp.eq}) and 
Eq.(\ref{eq.psi}), after integrating out the radial order parameter,  
GP equation reduced to one dimension reads,
\begin{equation}
 \left[ -\frac{\hbar^2}{2m}\frac{\partial ^2}{\partial z ^2} 
 + V^a(z) + u |\psi|^2 \right]\psi = i \hbar \frac{\partial \psi}{\partial t},
\label{1d.gp}
\end{equation}
where $u = 2 a\hbar\omega$ and $V^a(z) = \hbar\omega$ for $-L/2<z<L/2$ 
and zero otherwise. The order parameter satisfies normalization 
condition:
$ \int_{-L/2}^{L/2}dz |\psi(z)|^2 = N$.
It is well known, since the pioneering work of Zakharov and Shabat \cite{Zakharov}, 
that equation (\ref{1d.gp}) allows for grey solitons solutions. In terms of 
atoms per unit length along axial direction $\sigma(z) = |\psi(z)|^2$ and a 
dynamical phase, solution at $t=0$ is given by \cite{Jackson-02}
$\psi(z) = \sqrt{\sigma(z)} e^{i \phi (z)}$,
where 
\begin{equation} \no
 \sigma(z)  =  \sigma_0\left(1 - \frac{\cos^2(\theta)}{\cosh^2(z 
                 \cos(\theta)/\zeta)}\right)~~{\rm and}
\end{equation}
\begin{equation}
\phi(z) =  \tan^{-1}\left( \frac{c_w}{u}
                \sqrt{(1-u^2/c_w^2)}\tanh{\frac{\sqrt{(1-u^2/c_w^2)}z}{\zeta}}
                \right).
\label{soliton_phase} 
\end{equation}
Here velocity of soliton is given by $u$ and velocity of sound is $c_w$. 
Parameter $\theta$ is defined as $\theta = \sin^{-1}(u/c_w)$, whereas $\zeta$
 is defined as $\zeta = 2\xi(n_0)$,  and $\sigma_0 = n_0\pi a_{\rm osc}^2$ is 
the linear density far away from the soliton, \emph{i.e.,} in the domain 
$v\propto\phi'(z)\rightarrow0$. Notice that above soliton becomes `dark', 
\emph{i.e,} condensate density $\sigma(z)$ vanishes, at $z=0$ only when $u=0$. 
Therefore, by construction, a dark soliton in this model is static. Amazingly, 
in the limit $u$ approaches zero, from Eq.(\ref{soliton_phase}), one sees that 
the phase of the dark soliton behaves like a Heaviside step function with 
height $\pi$ and becomes discontinuous at $z=0$ \emph{,i.e.,} 
\begin{equation}\no
  \phi(z)=\left \{ \begin{aligned}
          &-\frac{\pi}{2} && \text{if }  z < 0, \\
          & 0            && \text{if }   z  = 0, \\
          & \frac{\pi}{2} &&\text{if }   z  > 0. 
       \end{aligned} \right.
\end{equation}
It is this discontinuity that makes the dark soliton a 1D counterpart of vortex. 
A cautious reader may ask that the above mentioned solitons are strictly solutions 
of Eq. (\ref{1d.gp}), when $\psi(z)$ is defined over whole of real line. In present case, because of particle-in-a-box type potential, it is only defined in domain $\frac{-L}{2} \leq z \leq \frac{L}{2}$. It was shown by Carr \emph{et. al.} 
\cite{Carr}, that dark soliton is indeed a genuine solution in this case too.
It can be easily shown that the energy required to 
create a dark soliton in otherwise homogeneous and uniform condensate is 
\cite{Jackson-02}
\begin{equation}\no
 \varepsilon = \frac{4\sqrt{2}}{3}\sqrt{(\sigma_0 a)}
               \sigma_0a_{\rm osc}\hbar\omega.
\end{equation}
In terms of $\sigma(z)$, the criterion for applicability of Eq.(\ref{1d.gp}), 
\emph{i.e.,} $2\pi a\int r|\Psi(r,z)|^2 dr\ll1$, becomes
$ \sigma(z) a \ll 1 $.
The transition between weakly interacting to strongly interacting high 
density limit occurs for $\sigma(z)a\sim0.25$ \cite{Jackson-98}. If $N$ is 
large ($N\rightarrow\infty$) such that $\xi/L\rightarrow0$, it can be argued 
that a dark soliton can be created without disturbing the other soliton 
\cite{Carr}. In reality, however, $N$ is finite, and $\xi/L$ is quite small 
but non-zero. It was shown in Refs. \cite{Zakharov,Carr}, and we have also 
checked numerically that above argument still holds as long as the distance between two dark solitons is much greater than the healing length. So the energy required to create $K$ solitons is 
$ \varepsilon_K = K \varepsilon$. Entropy associated with $K$ solitons is given by $K k \text{ln} W$, where $W$ is number of independent states soliton can occupy, and it is easy to see that $W = L/\xi$.  Hence, Helmholtz free energy for $K$ solitons is:
\begin{eqnarray}
 F & = & K \left( \frac{4\sqrt{2}}{3}\sqrt{(\sigma_0 a)}\sigma_0a_{\rm osc}
           \hbar \omega - kT\ln \frac{L}{\xi} \right),
\label{eq.free.energy}
\end{eqnarray} 
which changes sign at a characteristic temperature 
\begin{equation}
 T_{BKT} = \frac{4\sqrt{2}}{3k\ln (L/\xi)}\sqrt{(\sigma_0 a)}
           \sigma_0a_{\rm osc} \hbar\omega.
\label{eq.tc}
\end{equation}

\begin{figure}[h]
\begin{center}
  \includegraphics[width=8.5cm]{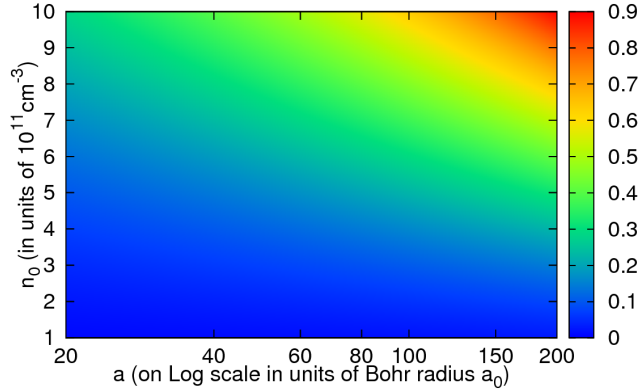}
  \caption{Density plot depicting $T_{BKT}$ (in nano Kelvins) as a function of bulk density $n_0$
           and s-wave scattering length $a$}
  \label{tbkt}
\end{center}
\end{figure}

So exactly like in 2D BEC, we see that in 1D BEC also there exists a 
finite critical temperature $T_{BKT}$, above which a uniform homogeneous 
condensate becomes thermodynamically unstable and spontaneous proliferation 
of dark solitons takes place. This indicates that 1D BEC exhibits a non-symmetry 
breaking BKT phase transition. Note that critical temperature obtained above, 
depends on various tunable parameters like $a$ and $\sigma_{0}$. In a typical 
cold atom experiment, the maximum density in trap ($n_0$) can vary from 
$10^{11}-10^{15}$cm$^{-3}$ \cite{Leggett}, whereas typically $\omega$ varies
from a few Hertz to kilo Hertz. For $\omega = 130 \,$ Hz, 
$L = 500\xi$, density around $10^{11}$cm$^{-3}$, and scattering length around $50 \, a_{0}$ ($a_{0}$ is Bohr radius), one finds that
$T_{BKT}$ lies in sub nano Kelvin range ($\sim 0.1 \, K$), which is an order of magnitude less than a typical BEC transition temperature. Figure (\ref{tbkt}) shows dependence of $T_{BKT}$ as a function of 
scattering length and density, and it is clear that it can be conveniently tuned by varying density and/or scattering length at will.

Dimensionally reduced GP equation (\ref{1d.gp}) can be obtained by 
extremizing the action $S = \int dxdt \, \mathscr{L}$, where
\begin{align} \no
\mathscr{L} =  i \hbar \psi^{\ast} \frac{\d \psi}{\d t} - 
\frac{\hbar^{2}}{2 m } 
\left| \frac{\d \psi}{\d z} \right|^{2} - {\psi}^{\ast} 
 \left( V^{a}(z) + u |\psi|^{2} 
\right) \psi.
\end{align}
In case when background condensate density is constant, only dynamical 
degree of freedom left is phase, since $\psi(x,t) = \sqrt{\sigma_{0}} 
e^{i \theta(x,t)}$. Dynamics of $\theta$ field is governed by Lagrangian
\begin{equation} \no 
\mathscr{L}_{\theta} = - \hbar \sigma_{0} \frac{\d \theta}{\d t} + 
\frac{\hbar^{2} \sigma_{0}}{2m} \theta \frac{\d^{2} \theta}{\d z^{2}}. 
\end{equation}

%

Provided that there are no dark solitons, static spatial phase correlation 
function can be easily found from the above Lagrangian and is
\begin{align} \no
\langle \theta(z) \theta(0) \rangle = \frac{m}{2 \hbar^{2}} |z|,
\end{align}   
which clearly indicates that system indeed has phase coherence, a 
characteristic feature of any BEC. However, situation changes when dark 
solitons are present. Phase field now can be decomposed into a regular 
continuous part and a discontinuous part, $i.e.,$ 
$\theta(z,t) = \theta_{reg}(z,t) + \theta_{dis}(z)$, respectively. 
Substituting this in the above Lagrangian, one finds
\begin{align} \no
\mathscr{L}_{\theta} =& - \hbar \sigma_{0} \frac{\d \theta_{reg}}{\d t} + 
\frac{\hbar^{2} \sigma_{0}}{2m} \theta_{reg} \frac{\d^{2} 
\theta_{reg}}{\d z^{2}} \\ \no &+ \frac{\hbar^{2} \sigma_{0}}{2m} 
\theta_{dis} \frac{\d^{2} \theta_{dis}}{\d z^{2}} + \frac{\hbar^{2} 
\sigma_{0}}{m} \theta_{dis} \frac{\d^{2} \theta_{reg}}{\d z^{2}}. 
\end{align}     
The effect of dark solitons on phase correlation can be inferred by first 
integrating out the singular part of field to yield an 
effective action for regular field. Very interestingly, one finds that 
integration of $\theta_{dis}$ gives rise to a term $- \frac{\hbar^{2} 
\sigma_{0}}{2m} \theta_{reg} \frac{\d^{2} \theta_{reg}}{\d z^{2}}$, 
which exactly cancels the gradient term already present in the action. Hence, 
phase field in presence of solitons does not propagate, and system losses 
long range phase correlation. In the absence of solitons, phase variation 
along the condensate profile is rather slow, and hence such a condensate would
yield a well defined interference pattern, an authentic signature of coherence.
However, when solitons are present, phase receives random discontinuous kicks 
along the profile, as a result phase ultimately gets averaged out. Such a 
condensate would certainly exhibit poor interference pattern, an indication of
loss of coherence. Hence we conclude that high temperature solitonic phase is actually
a disordered phase, and transition from uniform condensate to solitonic state 
is a phase transition.

We have shown that like 2D BEC, 1D BEC can also exhibit a BKT phase transition
at finite temperature. The role played by vortices in 2D is, in 1D case, 
played by dark solitons. It is found that critical temperature for this 
transition depends on atom number density and scattering length, both of 
which can be finely controlled in a typical experiment. It is this particular 
feature that could facilitate and strengthen possibility of an experimental 
realization of this transition. One may wonder whether above discussed phase transition 
indicates violation of Coleman-Mermin-Wagner-Hohenberg theorem \cite{Chaikin,Huang}. This theorem states that a continuous symmetry in given infinite system (not bounded by a finite volume), which is homogeneous and isotropic with short range local interactions, can not be broken at zero temperature in one spatial dimension and at finite temperature in two spatial dimensions. The proof of above theorem is
based on the fact that in such a system, breaking of a continuous symmetry 
results in existence of gapless Goldstone modes which by construction lead to infrared divergence. It was shown that in one spatial dimension, this divergence is so severe that it does not allow ordered phase to exist, and hence a continuous symmetry is not broken even at absolute zero temperature. As is
obvious, this theorem does not hold for our system of 1D BEC for several reasons. Firstly, we are
dealing with a system which is externally confined into a finite volume by trapping potentials, and
hence is neither homogeneous nor isotropic and certainly not of infinite extent. Further, we are actually looking at a quasi 1D BEC, where BEC itself lives in 3D but at sufficiently low energies, degrees of freedom along other dimensions are frozen. Moreover in BKT phase transition, no symmetry of Hamiltonian is spontaneously broken by ground state \cite{Chaikin,Huang,Nagaosa}. So we see that Coleman-Mermin-Wagner-Hohenberg theorem does not hold for our system. Possibility of BKT transition in 1D systems (in thermodynamic limit) have been studied in Refs. \cite{popov,afonin}, and both find transition temperature to be absolute zero. Interestingly, taking $L \rightarrow \infty$ in equation (\ref{eq.tc}), one sees that $T_{BKT}$ goes to zero, which agrees with their results. Issues pertaining to critical behavior and universality class of this transition are currently being investigated and will be published in due course.



\bibliography{kt.phase.transition.1d.bib}

\providecommand{\noopsort}[1]{}\providecommand{\singleletter}[1]{#1}%
\begin{thebibliography}{32}%
\makeatletter
\providecommand \@ifxundefined [1]{%
 \@ifx{#1\undefined}
}%
\providecommand \@ifnum [1]{%
 \ifnum #1\expandafter \@firstoftwo
 \else \expandafter \@secondoftwo
 \fi
}%
\providecommand \@ifx [1]{%
 \ifx #1\expandafter \@firstoftwo
 \else \expandafter \@secondoftwo
 \fi
}%
\providecommand \natexlab [1]{#1}%
\providecommand \enquote  [1]{``#1''}%
\providecommand \bibnamefont  [1]{#1}%
\providecommand \bibfnamefont [1]{#1}%
\providecommand \citenamefont [1]{#1}%
\providecommand \href@noop [0]{\@secondoftwo}%
\providecommand \href [0]{\begingroup \@sanitize@url \@href}%
\providecommand \@href[1]{\@@startlink{#1}\@@href}%
\providecommand \@@href[1]{\endgroup#1\@@endlink}%
\providecommand \@sanitize@url [0]{\catcode `\\12\catcode `\$12\catcode
  `\&12\catcode `\#12\catcode `\^12\catcode `\_12\catcode `\%12\relax}%
\providecommand \@@startlink[1]{}%
\providecommand \@@endlink[0]{}%
\providecommand \url  [0]{\begingroup\@sanitize@url \@url }%
\providecommand \@url [1]{\endgroup\@href {#1}{\urlprefix }}%
\providecommand \urlprefix  [0]{URL }%
\providecommand \Eprint [0]{\href }%
\providecommand \doibase [0]{http://dx.doi.org/}%
\providecommand \selectlanguage [0]{\@gobble}%
\providecommand \bibinfo  [0]{\@secondoftwo}%
\providecommand \bibfield  [0]{\@secondoftwo}%
\providecommand \translation [1]{[#1]}%
\providecommand \BibitemOpen [0]{}%
\providecommand \bibitemStop [0]{}%
\providecommand \bibitemNoStop [0]{.\EOS\space}%
\providecommand \EOS [0]{\spacefactor3000\relax}%
\providecommand \BibitemShut  [1]{\csname bibitem#1\endcsname}%
\let\auto@bib@innerbib\@empty
\bibitem [{\citenamefont {Chaikin}\ and\ \citenamefont
  {Lubensky}(2000)}]{Chaikin}%
  \BibitemOpen
  \bibfield  {author} {\bibinfo {author} {\bibfnamefont {P.~M.}\ \bibnamefont
  {Chaikin}}\ and\ \bibinfo {author} {\bibfnamefont {T.~C.}\ \bibnamefont
  {Lubensky}},\ }\href@noop {} {\emph {\bibinfo {title} {Principles of
  condensed matter physics}}}\ (\bibinfo  {publisher} {Cambridge University
  Press},\ \bibinfo {year} {2000})\BibitemShut {NoStop}%
\bibitem [{\citenamefont {Huang}(2010)}]{Huang}%
  \BibitemOpen
  \bibfield  {author} {\bibinfo {author} {\bibfnamefont {K.}~\bibnamefont
  {Huang}},\ }\href@noop {} {\emph {\bibinfo {title} {Quantum field theory:
  From operators to path integrals}}}\ (\bibinfo  {publisher} {Wiley-VCH},\
  \bibinfo {year} {2010})\BibitemShut {NoStop}%
\bibitem [{\citenamefont {MacKenzie}\ \emph {et~al.}(1993)\citenamefont
  {MacKenzie}, \citenamefont {Panigrahi},\ and\ \citenamefont
  {Sakhi}}]{Mackenzie1}%
  \BibitemOpen
  \bibfield  {author} {\bibinfo {author} {\bibfnamefont {R.}~\bibnamefont
  {MacKenzie}}, \bibinfo {author} {\bibfnamefont {P.}~\bibnamefont
  {Panigrahi}}, \ and\ \bibinfo {author} {\bibfnamefont {S.}~\bibnamefont
  {Sakhi}},\ }\href@noop {} {\bibfield  {journal} {\bibinfo  {journal} {Phys.
  Rev. B}\ }\textbf {\bibinfo {volume} {48}},\ \bibinfo {pages} {3892}
  (\bibinfo {year} {1993})}\BibitemShut {NoStop}%
\bibitem [{\citenamefont {MacKenzie}\ \emph {et~al.}(1994)\citenamefont
  {MacKenzie}, \citenamefont {Panigrahi},\ and\ \citenamefont
  {Sakhi}}]{Mackenzie2}%
  \BibitemOpen
  \bibfield  {author} {\bibinfo {author} {\bibfnamefont {R.}~\bibnamefont
  {MacKenzie}}, \bibinfo {author} {\bibfnamefont {P.}~\bibnamefont
  {Panigrahi}}, \ and\ \bibinfo {author} {\bibfnamefont {S.}~\bibnamefont
  {Sakhi}},\ }\href@noop {} {\bibfield  {journal} {\bibinfo  {journal}
  {International Journal of Modern Physics A}\ }\textbf {\bibinfo {volume}
  {9}},\ \bibinfo {pages} {3603} (\bibinfo {year} {1994})}\BibitemShut
  {NoStop}%
\bibitem [{\citenamefont {Nagaosa}(1999)}]{Nagaosa}%
  \BibitemOpen
  \bibfield  {author} {\bibinfo {author} {\bibfnamefont {N.}~\bibnamefont
  {Nagaosa}},\ }\href@noop {} {\emph {\bibinfo {title} {Quantum field theory in
  condensed matter physics}}}\ (\bibinfo  {publisher} {Springer Verlag},\
  \bibinfo {year} {1999})\BibitemShut {NoStop}%
\bibitem [{\citenamefont {Berezinskii}(1971)}]{Berezinskii}%
  \BibitemOpen
  \bibfield  {author} {\bibinfo {author} {\bibfnamefont {V.~L.}\ \bibnamefont
  {Berezinskii}},\ }\href@noop {} {\bibfield  {journal} {\bibinfo  {journal}
  {Soviet Journal of Experimental and Theoretical Physics}\ }\textbf {\bibinfo
  {volume} {32}},\ \bibinfo {pages} {493} (\bibinfo {year} {1971})}\BibitemShut
  {NoStop}%
\bibitem [{\citenamefont {Kosterlitz}\ and\ \citenamefont
  {Thouless}(1973)}]{Kosterlitz}%
  \BibitemOpen
  \bibfield  {author} {\bibinfo {author} {\bibfnamefont {J.~M.}\ \bibnamefont
  {Kosterlitz}}\ and\ \bibinfo {author} {\bibfnamefont {D.~J.}\ \bibnamefont
  {Thouless}},\ }\href {\doibase 10.1088/0022-3719/6/7/010} {\bibfield
  {journal} {\bibinfo  {journal} {J. Phys. C: Solid State Physics}\ }\textbf
  {\bibinfo {volume} {6}},\ \bibinfo {pages} {1181} (\bibinfo {year}
  {1973})}\BibitemShut {NoStop}%
\bibitem [{\citenamefont {Pethick}\ and\ \citenamefont
  {Smith}(2002)}]{Pethick}%
  \BibitemOpen
  \bibfield  {author} {\bibinfo {author} {\bibfnamefont {C.}~\bibnamefont
  {Pethick}}\ and\ \bibinfo {author} {\bibfnamefont {H.}~\bibnamefont
  {Smith}},\ }\href@noop {} {\emph {\bibinfo {title} {Bose-Einstein
  condensation in dilute gases}}}\ (\bibinfo  {publisher} {Cambridge University
  Press},\ \bibinfo {year} {2002})\BibitemShut {NoStop}%
\bibitem [{\citenamefont {Schweikhard}\ \emph {et~al.}(2007)\citenamefont
  {Schweikhard}, \citenamefont {Tung},\ and\ \citenamefont
  {Cornell}}]{Schweikhard}%
  \BibitemOpen
  \bibfield  {author} {\bibinfo {author} {\bibfnamefont {V.}~\bibnamefont
  {Schweikhard}}, \bibinfo {author} {\bibfnamefont {S.}~\bibnamefont {Tung}}, \
  and\ \bibinfo {author} {\bibfnamefont {E.~A.}\ \bibnamefont {Cornell}},\
  }\href {\doibase 10.1103/PhysRevLett.99.030401} {\bibfield  {journal}
  {\bibinfo  {journal} {Phys. Rev. Lett.}\ }\textbf {\bibinfo {volume} {99}},\
  \bibinfo {pages} {030401} (\bibinfo {year} {2007})}\BibitemShut {NoStop}%
\bibitem [{\citenamefont {Simula}\ \emph {et~al.}(2005)\citenamefont {Simula},
  \citenamefont {Lee},\ and\ \citenamefont {Hutchinson}}]{Simula-05}%
  \BibitemOpen
  \bibfield  {author} {\bibinfo {author} {\bibfnamefont {T.~P.}\ \bibnamefont
  {Simula}}, \bibinfo {author} {\bibfnamefont {M.~D.}\ \bibnamefont {Lee}}, \
  and\ \bibinfo {author} {\bibfnamefont {D.~A.~W.}\ \bibnamefont
  {Hutchinson}},\ }\href {\doibase 10.1080/09500830500256587} {\bibfield
  {journal} {\bibinfo  {journal} {Philosophical Magazine Letters}\ }\textbf
  {\bibinfo {volume} {85}},\ \bibinfo {pages} {395} (\bibinfo {year}
  {2005})}\BibitemShut {NoStop}%
\bibitem [{\citenamefont {Simula}\ and\ \citenamefont
  {Blakie}(2006)}]{Simula-06}%
  \BibitemOpen
  \bibfield  {author} {\bibinfo {author} {\bibfnamefont {T.~P.}\ \bibnamefont
  {Simula}}\ and\ \bibinfo {author} {\bibfnamefont {P.~B.}\ \bibnamefont
  {Blakie}},\ }\href {\doibase 10.1103/PhysRevLett.96.020404} {\bibfield
  {journal} {\bibinfo  {journal} {Phys. Rev. Lett.}\ }\textbf {\bibinfo
  {volume} {96}},\ \bibinfo {pages} {020404} (\bibinfo {year}
  {2006})}\BibitemShut {NoStop}%
\bibitem [{\citenamefont {Hadzibabic}\ \emph {et~al.}(2006)\citenamefont
  {Hadzibabic}, \citenamefont {Kruger}, \citenamefont {Cheneau}, \citenamefont
  {Battelier},\ and\ \citenamefont {Dalibard}}]{Hadzibabic}%
  \BibitemOpen
  \bibfield  {author} {\bibinfo {author} {\bibfnamefont {Z.}~\bibnamefont
  {Hadzibabic}}, \bibinfo {author} {\bibfnamefont {P.}~\bibnamefont {Kruger}},
  \bibinfo {author} {\bibfnamefont {M.}~\bibnamefont {Cheneau}}, \bibinfo
  {author} {\bibfnamefont {B.}~\bibnamefont {Battelier}}, \ and\ \bibinfo
  {author} {\bibfnamefont {J.}~\bibnamefont {Dalibard}},\ }\href {\doibase
  10.1038/nature04851} {\bibfield  {journal} {\bibinfo  {journal} {Nature}\
  }\textbf {\bibinfo {volume} {441}},\ \bibinfo {pages} {1118} (\bibinfo {year}
  {2006})}\BibitemShut {NoStop}%
\bibitem [{\citenamefont {Susanto}\ \emph {et~al.}(2007)\citenamefont
  {Susanto}, \citenamefont {Kevrekidis}, \citenamefont {Carretero-Gonz\'alez},
  \citenamefont {Malomed}, \citenamefont {Frantzeskakis},\ and\ \citenamefont
  {Bishop}}]{Susanto}%
  \BibitemOpen
  \bibfield  {author} {\bibinfo {author} {\bibfnamefont {H.}~\bibnamefont
  {Susanto}}, \bibinfo {author} {\bibfnamefont {P.~G.}\ \bibnamefont
  {Kevrekidis}}, \bibinfo {author} {\bibfnamefont {R.}~\bibnamefont
  {Carretero-Gonz\'alez}}, \bibinfo {author} {\bibfnamefont {B.~A.}\
  \bibnamefont {Malomed}}, \bibinfo {author} {\bibfnamefont {D.~J.}\
  \bibnamefont {Frantzeskakis}}, \ and\ \bibinfo {author} {\bibfnamefont
  {A.~R.}\ \bibnamefont {Bishop}},\ }\href {\doibase
  10.1103/PhysRevA.75.055601} {\bibfield  {journal} {\bibinfo  {journal} {Phys.
  Rev. A}\ }\textbf {\bibinfo {volume} {75}},\ \bibinfo {pages} {055601}
  (\bibinfo {year} {2007})}\BibitemShut {NoStop}%
\bibitem [{\citenamefont {Rodrigues}\ \emph {et~al.}(2009)\citenamefont
  {Rodrigues}, \citenamefont {Kevrekidis}, \citenamefont
  {Carretero-Gonz\'alez}, \citenamefont {Frantzeskakis}, \citenamefont
  {Schmelcher}, \citenamefont {Alexander},\ and\ \citenamefont
  {Kivshar}}]{Rodrigues}%
  \BibitemOpen
  \bibfield  {author} {\bibinfo {author} {\bibfnamefont {A.~S.}\ \bibnamefont
  {Rodrigues}}, \bibinfo {author} {\bibfnamefont {P.~G.}\ \bibnamefont
  {Kevrekidis}}, \bibinfo {author} {\bibfnamefont {R.}~\bibnamefont
  {Carretero-Gonz\'alez}}, \bibinfo {author} {\bibfnamefont {D.~J.}\
  \bibnamefont {Frantzeskakis}}, \bibinfo {author} {\bibfnamefont
  {P.}~\bibnamefont {Schmelcher}}, \bibinfo {author} {\bibfnamefont {T.~J.}\
  \bibnamefont {Alexander}}, \ and\ \bibinfo {author} {\bibfnamefont {Y.~S.}\
  \bibnamefont {Kivshar}},\ }\href {\doibase 10.1103/PhysRevA.79.043603}
  {\bibfield  {journal} {\bibinfo  {journal} {Phys. Rev. A}\ }\textbf {\bibinfo
  {volume} {79}},\ \bibinfo {pages} {043603} (\bibinfo {year}
  {2009})}\BibitemShut {NoStop}%
\bibitem [{\citenamefont {Gladush}\ \emph {et~al.}(2009)\citenamefont
  {Gladush}, \citenamefont {Kamchatnov}, \citenamefont {Shi}, \citenamefont
  {Kevrekidis}, \citenamefont {Frantzeskakis},\ and\ \citenamefont
  {Malomed}}]{Gladush}%
  \BibitemOpen
  \bibfield  {author} {\bibinfo {author} {\bibfnamefont {Y.~G.}\ \bibnamefont
  {Gladush}}, \bibinfo {author} {\bibfnamefont {A.~M.}\ \bibnamefont
  {Kamchatnov}}, \bibinfo {author} {\bibfnamefont {Z.}~\bibnamefont {Shi}},
  \bibinfo {author} {\bibfnamefont {P.~G.}\ \bibnamefont {Kevrekidis}},
  \bibinfo {author} {\bibfnamefont {D.~J.}\ \bibnamefont {Frantzeskakis}}, \
  and\ \bibinfo {author} {\bibfnamefont {B.~A.}\ \bibnamefont {Malomed}},\
  }\href {\doibase 10.1103/PhysRevA.79.033623} {\bibfield  {journal} {\bibinfo
  {journal} {Phys. Rev. A}\ }\textbf {\bibinfo {volume} {79}},\ \bibinfo
  {pages} {033623} (\bibinfo {year} {2009})}\BibitemShut {NoStop}%
\bibitem [{\citenamefont {Sasaki}\ \emph {et~al.}(2010)\citenamefont {Sasaki},
  \citenamefont {Suzuki},\ and\ \citenamefont {Saito}}]{Sasaki}%
  \BibitemOpen
  \bibfield  {author} {\bibinfo {author} {\bibfnamefont {K.}~\bibnamefont
  {Sasaki}}, \bibinfo {author} {\bibfnamefont {N.}~\bibnamefont {Suzuki}}, \
  and\ \bibinfo {author} {\bibfnamefont {H.}~\bibnamefont {Saito}},\ }\href
  {\doibase 10.1103/PhysRevLett.104.150404} {\bibfield  {journal} {\bibinfo
  {journal} {Phys. Rev. Lett.}\ }\textbf {\bibinfo {volume} {104}},\ \bibinfo
  {pages} {150404} (\bibinfo {year} {2010})}\BibitemShut {NoStop}%
\bibitem [{\citenamefont {Neely}\ \emph {et~al.}(2010)\citenamefont {Neely},
  \citenamefont {Samson}, \citenamefont {Bradley}, \citenamefont {Davis},\ and\
  \citenamefont {Anderson}}]{Neely}%
  \BibitemOpen
  \bibfield  {author} {\bibinfo {author} {\bibfnamefont {T.~W.}\ \bibnamefont
  {Neely}}, \bibinfo {author} {\bibfnamefont {E.~C.}\ \bibnamefont {Samson}},
  \bibinfo {author} {\bibfnamefont {A.~S.}\ \bibnamefont {Bradley}}, \bibinfo
  {author} {\bibfnamefont {M.~J.}\ \bibnamefont {Davis}}, \ and\ \bibinfo
  {author} {\bibfnamefont {B.~P.}\ \bibnamefont {Anderson}},\ }\href {\doibase
  10.1103/PhysRevLett.104.160401} {\bibfield  {journal} {\bibinfo  {journal}
  {Phys. Rev. Lett.}\ }\textbf {\bibinfo {volume} {104}},\ \bibinfo {pages}
  {160401} (\bibinfo {year} {2010})}\BibitemShut {NoStop}%
\bibitem [{\citenamefont {Freilich}\ \emph {et~al.}(2010)\citenamefont
  {Freilich}, \citenamefont {Bianchi}, \citenamefont {Kaufman}, \citenamefont
  {Langin},\ and\ \citenamefont {Hall}}]{Freilich}%
  \BibitemOpen
  \bibfield  {author} {\bibinfo {author} {\bibfnamefont {D.~V.}\ \bibnamefont
  {Freilich}}, \bibinfo {author} {\bibfnamefont {D.~M.}\ \bibnamefont
  {Bianchi}}, \bibinfo {author} {\bibfnamefont {A.~M.}\ \bibnamefont
  {Kaufman}}, \bibinfo {author} {\bibfnamefont {T.~K.}\ \bibnamefont {Langin}},
  \ and\ \bibinfo {author} {\bibfnamefont {D.~S.}\ \bibnamefont {Hall}},\
  }\href {\doibase 10.1126/science.1191224} {\bibfield  {journal} {\bibinfo
  {journal} {Science}\ }\textbf {\bibinfo {volume} {329}},\ \bibinfo {pages}
  {1182} (\bibinfo {year} {2010})}\BibitemShut {NoStop}%
\bibitem [{\citenamefont {Theocharis}\ \emph {et~al.}(2003)\citenamefont
  {Theocharis}, \citenamefont {Frantzeskakis}, \citenamefont {Kevrekidis},
  \citenamefont {Malomed},\ and\ \citenamefont {Kivshar}}]{Theocharis}%
  \BibitemOpen
  \bibfield  {author} {\bibinfo {author} {\bibfnamefont {G.}~\bibnamefont
  {Theocharis}}, \bibinfo {author} {\bibfnamefont {D.~J.}\ \bibnamefont
  {Frantzeskakis}}, \bibinfo {author} {\bibfnamefont {P.~G.}\ \bibnamefont
  {Kevrekidis}}, \bibinfo {author} {\bibfnamefont {B.~A.}\ \bibnamefont
  {Malomed}}, \ and\ \bibinfo {author} {\bibfnamefont {Y.~S.}\ \bibnamefont
  {Kivshar}},\ }\href {\doibase 10.1103/PhysRevLett.90.120403} {\bibfield
  {journal} {\bibinfo  {journal} {Phys. Rev. Lett.}\ }\textbf {\bibinfo
  {volume} {90}},\ \bibinfo {pages} {120403} (\bibinfo {year}
  {2003})}\BibitemShut {NoStop}%
\bibitem [{\citenamefont {Anderson}\ \emph {et~al.}(2001)\citenamefont
  {Anderson}, \citenamefont {Haljan}, \citenamefont {Regal}, \citenamefont
  {Feder}, \citenamefont {Collins}, \citenamefont {Clark},\ and\ \citenamefont
  {Cornell}}]{Anderson}%
  \BibitemOpen
  \bibfield  {author} {\bibinfo {author} {\bibfnamefont {B.~P.}\ \bibnamefont
  {Anderson}}, \bibinfo {author} {\bibfnamefont {P.~C.}\ \bibnamefont
  {Haljan}}, \bibinfo {author} {\bibfnamefont {C.~A.}\ \bibnamefont {Regal}},
  \bibinfo {author} {\bibfnamefont {D.~L.}\ \bibnamefont {Feder}}, \bibinfo
  {author} {\bibfnamefont {L.~A.}\ \bibnamefont {Collins}}, \bibinfo {author}
  {\bibfnamefont {C.~W.}\ \bibnamefont {Clark}}, \ and\ \bibinfo {author}
  {\bibfnamefont {E.~A.}\ \bibnamefont {Cornell}},\ }\href {\doibase
  10.1103/PhysRevLett.86.2926} {\bibfield  {journal} {\bibinfo  {journal}
  {Phys. Rev. Lett.}\ }\textbf {\bibinfo {volume} {86}},\ \bibinfo {pages}
  {2926} (\bibinfo {year} {2001})}\BibitemShut {NoStop}%
\bibitem [{\citenamefont {Zachary~Dutton}\ \emph {et~al.}(2001)\citenamefont
  {Zachary~Dutton}, \citenamefont {Budde}, \citenamefont {Slowe},\ and\
  \citenamefont {Hau}}]{Dutton}%
  \BibitemOpen
  \bibfield  {author} {\bibinfo {author} {\bibfnamefont {Z.}~\bibnamefont
  {Zachary~Dutton}}, \bibinfo {author} {\bibfnamefont {M.}~\bibnamefont
  {Budde}}, \bibinfo {author} {\bibfnamefont {C.}~\bibnamefont {Slowe}}, \ and\
  \bibinfo {author} {\bibfnamefont {L.~V.}\ \bibnamefont {Hau}},\ }\href
  {\doibase 10.1126/science.1062527} {\bibfield  {journal} {\bibinfo  {journal}
  {Science}\ }\textbf {\bibinfo {volume} {293}},\ \bibinfo {pages} {663}
  (\bibinfo {year} {2001})}\BibitemShut {NoStop}%
\bibitem [{\citenamefont {Shenoy}(1990)}]{Shenoy}%
  \BibitemOpen
  \bibfield  {author} {\bibinfo {author} {\bibfnamefont {S.~R.}\ \bibnamefont
  {Shenoy}},\ }\href {\doibase 10.1103/PhysRevB.42.8595} {\bibfield  {journal}
  {\bibinfo  {journal} {Phys. Rev. B}\ }\textbf {\bibinfo {volume} {42}},\
  \bibinfo {pages} {8595} (\bibinfo {year} {1990})}\BibitemShut {NoStop}%
\bibitem [{\citenamefont {Kleinert}(2008)}]{Kleinert}%
  \BibitemOpen
  \bibfield  {author} {\bibinfo {author} {\bibfnamefont {H.}~\bibnamefont
  {Kleinert}},\ }\href@noop {} {\emph {\bibinfo {title} {Multivalued Fields in
  Condensed Matter, Electrodynamics and Gravitation}}}\ (\bibinfo  {publisher}
  {World Scientific, Singapore},\ \bibinfo {year} {2008})\BibitemShut {NoStop}%
\bibitem [{\citenamefont {Carr}\ \emph {et~al.}(2000)\citenamefont {Carr},
  \citenamefont {Clark},\ and\ \citenamefont {Reinhardt}}]{Carr}%
  \BibitemOpen
  \bibfield  {author} {\bibinfo {author} {\bibfnamefont {L.~D.}\ \bibnamefont
  {Carr}}, \bibinfo {author} {\bibfnamefont {C.~W.}\ \bibnamefont {Clark}}, \
  and\ \bibinfo {author} {\bibfnamefont {W.~P.}\ \bibnamefont {Reinhardt}},\
  }\href {\doibase 10.1103/PhysRevA.62.063610} {\bibfield  {journal} {\bibinfo
  {journal} {Phys. Rev. A}\ }\textbf {\bibinfo {volume} {62}},\ \bibinfo
  {pages} {063610} (\bibinfo {year} {2000})}\BibitemShut {NoStop}%
\bibitem [{\citenamefont {Frantzeskakis}(2010)}]{Frantzeskakis}%
  \BibitemOpen
  \bibfield  {author} {\bibinfo {author} {\bibfnamefont {D.~J.}\ \bibnamefont
  {Frantzeskakis}},\ }\href {\doibase 10.1088/1751-8113/43/21/213001}
  {\bibfield  {journal} {\bibinfo  {journal} {J. Phys. A}\ }\textbf {\bibinfo
  {volume} {43}},\ \bibinfo {pages} {213001} (\bibinfo {year}
  {2010})}\BibitemShut {NoStop}%
\bibitem [{\citenamefont {Salasnich}\ \emph {et~al.}(2002)\citenamefont
  {Salasnich}, \citenamefont {Parola},\ and\ \citenamefont
  {Reatto}}]{Salasnich}%
  \BibitemOpen
  \bibfield  {author} {\bibinfo {author} {\bibfnamefont {L.}~\bibnamefont
  {Salasnich}}, \bibinfo {author} {\bibfnamefont {A.}~\bibnamefont {Parola}}, \
  and\ \bibinfo {author} {\bibfnamefont {L.}~\bibnamefont {Reatto}},\ }\href
  {\doibase 10.1103/PhysRevA.65.043614} {\bibfield  {journal} {\bibinfo
  {journal} {Phys. Rev. A}\ }\textbf {\bibinfo {volume} {65}},\ \bibinfo
  {pages} {043614} (\bibinfo {year} {2002})}\BibitemShut {NoStop}%
\bibitem [{\citenamefont {Zakharov}\ and\ \citenamefont
  {Shabat}(1973)}]{Zakharov}%
  \BibitemOpen
  \bibfield  {author} {\bibinfo {author} {\bibfnamefont {V.~E.}\ \bibnamefont
  {Zakharov}}\ and\ \bibinfo {author} {\bibfnamefont {A.~B.}\ \bibnamefont
  {Shabat}},\ }\href@noop {} {\bibfield  {journal} {\bibinfo  {journal} {Soviet
  Journal of Experimental and Theoretical Physics}\ }\textbf {\bibinfo {volume}
  {37}},\ \bibinfo {pages} {823} (\bibinfo {year} {1973})}\BibitemShut
  {NoStop}%
\bibitem [{\citenamefont {Jackson}\ and\ \citenamefont
  {Kavoulakis}(2002)}]{Jackson-02}%
  \BibitemOpen
  \bibfield  {author} {\bibinfo {author} {\bibfnamefont {A.~D.}\ \bibnamefont
  {Jackson}}\ and\ \bibinfo {author} {\bibfnamefont {G.~M.}\ \bibnamefont
  {Kavoulakis}},\ }\href {\doibase 10.1103/PhysRevLett.89.070403} {\bibfield
  {journal} {\bibinfo  {journal} {Phys. Rev. Lett.}\ }\textbf {\bibinfo
  {volume} {89}},\ \bibinfo {pages} {070403} (\bibinfo {year}
  {2002})}\BibitemShut {NoStop}%
\bibitem [{\citenamefont {Jackson}\ \emph {et~al.}(1998)\citenamefont
  {Jackson}, \citenamefont {Kavoulakis},\ and\ \citenamefont
  {Pethick}}]{Jackson-98}%
  \BibitemOpen
  \bibfield  {author} {\bibinfo {author} {\bibfnamefont {A.~D.}\ \bibnamefont
  {Jackson}}, \bibinfo {author} {\bibfnamefont {G.~M.}\ \bibnamefont
  {Kavoulakis}}, \ and\ \bibinfo {author} {\bibfnamefont {C.~J.}\ \bibnamefont
  {Pethick}},\ }\href {\doibase 10.1103/PhysRevA.58.2417} {\bibfield  {journal}
  {\bibinfo  {journal} {Phys. Rev. A}\ }\textbf {\bibinfo {volume} {58}},\
  \bibinfo {pages} {2417} (\bibinfo {year} {1998})}\BibitemShut {NoStop}%
\bibitem [{\citenamefont {Leggett}(2001)}]{Leggett}%
  \BibitemOpen
  \bibfield  {author} {\bibinfo {author} {\bibfnamefont {A.~J.}\ \bibnamefont
  {Leggett}},\ }\href {\doibase 10.1103/RevModPhys.73.307} {\bibfield
  {journal} {\bibinfo  {journal} {Rev. Mod. Phys.}\ }\textbf {\bibinfo {volume}
  {73}},\ \bibinfo {pages} {307} (\bibinfo {year} {2001})}\BibitemShut
  {NoStop}%
\bibitem [{\citenamefont {Popov}\ and\ \citenamefont {Niederle}(2001)}]{popov}%
  \BibitemOpen
  \bibfield  {author} {\bibinfo {author} {\bibfnamefont {V.}~\bibnamefont
  {Popov}}\ and\ \bibinfo {author} {\bibfnamefont {J.}~\bibnamefont
  {Niederle}},\ }\href@noop {} {\emph {\bibinfo {title} {Functional integrals
  in quantum field theory and statistical physics}}},\ Vol.~\bibinfo {volume}
  {8}\ (\bibinfo  {publisher} {Kluwer Academic},\ \bibinfo {year}
  {2001})\BibitemShut {NoStop}%
\bibitem [{\citenamefont {Afonin}\ and\ \citenamefont {Petrov}(2008)}]{afonin}%
  \BibitemOpen
  \bibfield  {author} {\bibinfo {author} {\bibfnamefont {V.}~\bibnamefont
  {Afonin}}\ and\ \bibinfo {author} {\bibfnamefont {V.}~\bibnamefont
  {Petrov}},\ }\href@noop {} {\bibfield  {journal} {\bibinfo  {journal}
  {Journal of Experimental and Theoretical Physics}\ }\textbf {\bibinfo
  {volume} {107}},\ \bibinfo {pages} {542} (\bibinfo {year}
  {2008})}\BibitemShut {NoStop}%
\end{thebibliography}%

%
%
\end{document}